# Comment on Pisarenko et al. "Characterization of the Tail of the Distribution of Earthquake Magnitudes by Combining the GEV and GPD Descriptions of Extreme Value Theory"

Mathias Raschke

Institution: freelancer (free researcher)

Adress: Stolze-Schrey-Str.1 , 65195 Wiesbaden, Germany

Tel.: 0049 611 98819561

Email: mathiasraschke@t-online.de

Abstract: In this short note, I comment on the research of Pisarenko et al. (2014) regarding the extreme value theory and statistics in case of earthquake magnitudes. The link between the generalized extreme value distribution (GEVD) as an asymptotic model for the block maxima of a random variable and the generalized Pareto distribution (GPD) as a model for the peak over thresholds (POT) of the same random variable is presented more clearly. Pisarenko et al. (2014) have inappropriately neglected that the approximations by GEVD and GPD work only asymptotically in most cases. This applies particularly for the truncated exponential distribution (TED), being a popular distribution model for earthquake magnitudes. I explain why the classical models and methods of the extreme value theory and statistics do not work well for truncated exponential distributions. As a consequence, the classical methods should be used for the estimation of the upper bound magnitude and corresponding parameters. Furthermore, different issues of statistical inference of Pisarenko et al. (2014) are commented on and alternatives are proposed. Arguments are presented why GPD and GEVD would work for different types of stochastic earthquake processes in time, and not only for the homogeneous (stationery) Poisson process as assumed by Pisarenko et al. (2014). The crucial point of earthquake magnitudes is the poor convergence of their tail distribution to the GPD, and not the earthquake process in time.

# 1. Introduction

This short note is a comment on the paper „Characterization of the Tail of the Distribution of Earthquake Magnitudes by Combining the GEV and GPD Descriptions of Extreme Value Theory" by Pisarenko et al. (2014, referred to as Pisarenko et al. in this note). Being a continuation of the research of Pisarenko et al. (2008), the authors suggest applying both the generalized extreme value distribution (GEVD) and the generalized Pareto distribution (GPD) to the distribution of extreme magnitudes for estimating the upper bound magnitude and the quantiles of the maximum magnitude of a defined time period. They also present the cumulative distribution function (CDF) of the truncated exponential distribution (TED) as a distribution model for earthquake magnitudes (see e.g. Cosentino et al., 1977). It is popular and often applied in the hazard models, e.g., for the Euro-Mediterranean region (SHARE, 2014). The CDF of the TED is written as

$$F(x) = \begin{array}{ll} 0; & x < \mathrm{m}_{min}; \\ \frac{1-\exp(-\beta(x-\mathrm{m}_{min}))}{1-\exp(-\beta(\mathrm{m}_{max}-\mathrm{m}_{min}))}; & \mathrm{m}_{min} \leq \mathrm{x} \leq \mathrm{m}_{max}; \\ 1; & x > \mathrm{m}_{max}. \end{array} \tag{1}$$

the exponential function is preferred here rather than the power function with base 10 as the exponential function is usually used in mathematics (e.g. Hannon and Dahiya, 1999). The parameter $m_{max}$ represents the upper bound magnitude and its estimation has been the subject of many studies (Pisarenko et al., 1996; Kijko and Graham, 1998; Raschke, 2012).

As aforementioned, Pisarenko et al. suggest applying the GEVD and the GPD for the estimation of the upper bound magnitude and explain the link between GEVD and GPD. I present this link in a more straightforward and transparent manner in the following section. Therein, I also explain the reasons why these models and methods of the extreme value statistics do not work well in the case of the TED and other truncated exponential distributions. In section 3, different issues concerning the parameter estimation for GPD and GEVD by Pisarenko et al. are commented on.

Additionally, Pisarenko et al. decluster the earthquake catalogs and only consider the main events before applying the extreme value analysis to the empirical earthquake data. This is based on their assumption that the event occurrence has to follow a homogeneous Poisson process. This is, in fact, not necessary for applying extreme value models, as I will explain in section 4. Finally, the comments are summarized in section 5.

The notations of the extreme value theory and statistics mentioned in the following chapters are provided in Table A1 in the appendix, which also includes the corresponding symbols of Pisarenko et al..

## 2. Comments on the extreme value theory

The first important results of the extreme value theory have been achieved by Fischer and Tippett (1928) and Gnedenko (1943). Nowadays, the extreme value theory is a well-established field within probability theory and mathematical statistics. There are numerous studies available, dealing with many aspects in this field (e.g. Leadbetter et al., 1983; de Haan and Ferreira, 2006; Falk et al., 2011). One fundamental subject of the extreme value theory is the distribution of peaks over threshold (POT, see e.g. Coles, 2001, ch. 4; Berilant et al., 2004a and b, sec. 5.3), which is defined as

$$Y = X - x_{threshold}, \ Y \geq 0, \tag{2}$$

with a real-valued random variable $X$ and the excess $Y$ over a certain threshold. The threshold acts in the same way as $m_{min}$ in Eq. (1). Under certain conditions, the CDF of $Y$ increasingly approximates the GPD when the threshold is increasing. The CDF of the GPD is written as

$$H(x) = \begin{array}{ll} 1 - (1 + \gamma x/\sigma^*)^{-\frac{1}{\gamma}}, & \gamma \neq 0, x > 0, and\ x < -\sigma^*/\gamma\ \ if\ \ \gamma < 0 \\ 1 - exp(-x/\sigma^*) & \gamma = 0,\ \ x > 0, \end{array} \tag{3}$$

where the $\sigma^*$ is the scaling parameter and $\gamma$ is the extreme value index (also called the tail index). With the Weibull case the finite right end point is $\gamma$<0, with the Gumbel case it is $\gamma$=0, and with the Fréchet case is $\gamma$>0. The Gumbel case also represents the exponential distribution (ED). Therein, the scale parameter $\sigma^*$ is equivalent to the reciprocal scale parameter $1/\beta$ of the TED of Eq. (1) with an infinite $m_{max}$.

If the tail of a distribution is a member of one of the domains of attraction of the GPD, then the CDF of the block maxima

$$Z = max\{X_1, X_2, \ldots, X_n\} \tag{4}$$

can be approximated by the GEVD in case of a large block (sample) size $n$ (see e.g. Beirlant et al., 2004a and b, sec. 5.1), with CDF

$$G(x) = \begin{array}{ll} exp\left(-(1 + \gamma(x - \mu)/\sigma)^{-\frac{1}{\gamma}}\right), & \gamma \neq 0, x > \mu - \frac{\sigma}{\gamma}\ if\ \gamma > 0\ otherwise\ x < \mu - \frac{\sigma}{\gamma} \\ exp\left(-exp\left(-\frac{x-\mu}{\sigma}\right)\right), & \gamma = 0 \end{array} \tag{5}$$

The block size can also be determined by a defined length of time period of observation. The actual and exact distribution of the block maximum $Z$ is simply formulated for independent and identical distributed random variables $X$ with

$$G(x) = F(x)^n. \tag{6}$$

The probability that the maximum $Z \leq x$ is equal to the probability of no realization with $X > x$ in the block (sample). In case of a large $n$, we can apply the Poisson approximation (see e.g. Falk et al. 2011, Part I). This means that the number of realization $X > x$ is binomial distributed, which can be approximated by the Poisson distribution and can be written as

$$G(x) = Pr\{m = 0 | X > x\} = \exp(n(1 - F(x)). \tag{7}$$

Furthermore, $n$ and $F(x)$ can be replaced by $H(x)$ and the (average) number of excesses $n_{threshold}$ in case of a large sample size $n$. Hence, we get (bounds of $x$ according to Eq. (5,9))

$$G(x) = \begin{array}{l} exp\left(-n_{threshold}(1 + \gamma(x - x_{threshold})/\sigma^*)^{-\frac{1}{\gamma}}\right), \quad \gamma \neq 0 \\ \exp\left(-n_{threshold} exp\left(-\frac{x - x_{threshold}}{\sigma^*}\right)\right), \qquad \gamma = 0 \end{array} . \tag{8}$$

This equation is equivalent to Eq. (5) with the following parameter transformation

$$\sigma = n_{threshold}^{\gamma} \sigma^*, \tag{9a}$$

$$\mu = x_{threshold} - \sigma^*(1 - n_{threshold}^{\gamma})/\gamma \; if \; \gamma \neq 0, \tag{9b}$$

$$\mu = x_{threshold} + \sigma^* \ln(n_{threshold}) \; if \; \gamma = 0, \tag{9c}$$

and the extreme value index $\gamma$ is the same. This transformation is already well known and need no further explanation (see e.g. Coles, 2001, ch. 4). Note that a distribution, which is a member of one domain of attraction of the GEVD and GPD, has only one exact asymptotic extreme value index $\gamma$, although different estimations can be obtained.

A crucial point is the convergence speed. How fast does the GEVD approximate the distribution of block maxima in Eq. (6) and the GPD the actual tail of the distribution? For example, the exponential distribution is equal to the GPD with an extreme value index of $\gamma = 0$ and the tail of an ED is again an ED. The convergence speed is infinite. Correspondingly,

Leadbetter et al. (1983) have shown that the block maxima of an exponential distributed random variable converge to a GEVD with an extreme value index of $\gamma=0$. For applying the GEVD, no large size of the block maxima is needed (Beirlant et al., 2004, Fig. 2.9).

The TED is similar to the exponential distribution when the upper bound $m_{max}$ is relatively high. This applies in many cases, such as the magnitude distribution according to the well-known Gutenberg-Richter law. Nevertheless, the extreme value index of the TED is $\gamma=-1$ according to Leadbetter et al. (1983). This means that the POT of a TED converges to a uniform distribution. But the convergence is often poor due to the similarity between ED and TED. In Figure 1, the POTs of different TEDs with parameter $x_{threshold}=m_{min}$ of Eq. (1,2) and the corresponding GPD with an extreme value index of $\gamma=-1$ is presented. The scale parameter of the GPD is determined by $\sigma^*=m_{max}$. It is obvious that different TEDs with equal upper bounds have the same asymptotic tail distribution. But the approximation of the TEDs by a GPD works only for very small differences of $m_{max}-m_{min}$ in relation to the parameter $\beta$ ($\beta\approx2.3$ in case of earthquake magnitudes). Similarly, the block size has to be very large until the approximation of Eq. (6) with the GEVD works for the TED. This is in line with previous results by Raschke (2012, sec. 2.6), wherein the upper bound $m_{max}$ can be better estimated by the methods of Pisarenko et al. (1996), Kijko and Graham (1998) and Raschke (2012) than by extreme value statistics.

Of course, there are alternative distribution models for the magnitudes than the TED. The generalized truncated exponential distribution (GTED) is such one and has been formulated recently by Raschke (2014). In all cases, the magnitude distribution is similar to the ED (being a GPD with an extreme value index of $\gamma=0$) in a large share of the definition range of the random variable (cf. Gutenberg-Richter law). This implies a poor convergence of the upper tail of the magnitude distribution to the corresponding asymptotic GPD with an extreme value index of $\gamma \ll 0$ and hence I strongly advise against applying classical extreme value statistics for the approximation of earthquake magnitudes.

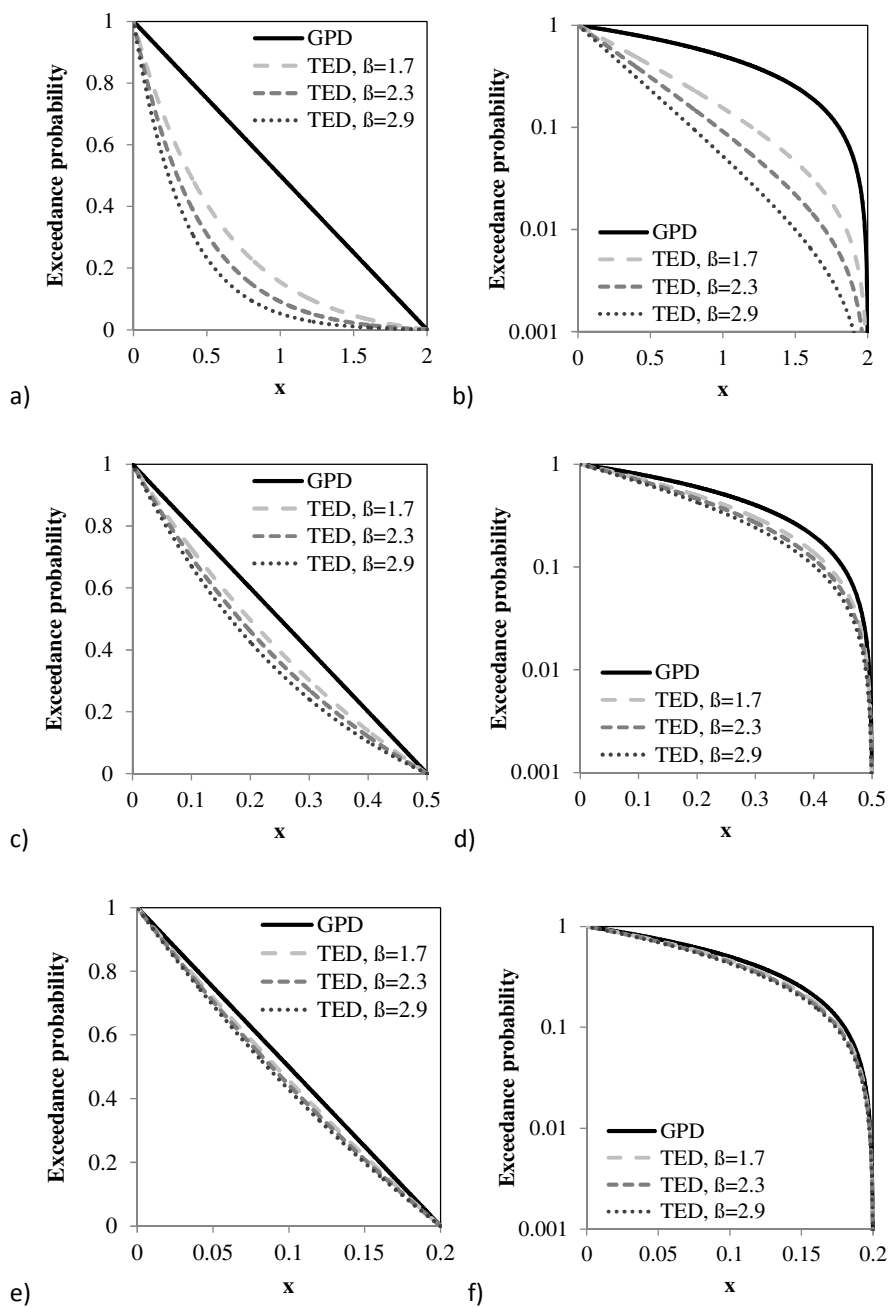

Fig.1: Demonstration of the poor convergence speed of the tail of the TED to the GPD for different scale parameters $\beta$ of the TED und upper bounds $m_{max}$, $m$=0. In all cases, the GPDs have an extreme value index of $\gamma$=-1 and the same upper bound as the TEDs: a) $m_{max}$=2, b) as a) but with logarithmized scale, c) $m_{max}$=0.5, d) as c) but with logarithmized scale, e) $m_{max}$=0.2, f) as e) but with logarithmized scale.

### 3. Comments on the statistical inference

Pisarenko et al. apply the GEVD using the method of block maxima and the GPD using the POT method to estimate the upper bound magnitude and the quantile of the maximum magnitude for a defined time period. In both cases, they use confusing procedures, which are inconsistent and differ from the classical extreme value statistics. Pisarenko et al. consider different lengths $T$ of the time periods for the method of block maxima. This is unusual but possible and is oriented towards the classical POT with different thresholds. The authors apply the moment estimation for the parameter estimation of the GEVD and refer to Pisarenko et al. (2008), which is in contrast to the usually used extreme value statistics, as e.g. presented by Cloes (2011, section 3.3) and Beirlant et al. (2004a and b, section 5.1). According to these references, the probability weighted moment (PWM) method and the maximum likelihood (ML) method should have been used by Pisarenko et al.. The authors justify their choice by pointing out the better estimation results in Pisarenko et al. (2008). But this argument is weak, because Pisarenko et al. (2008) did not investigate the asymptotic behavior of the moment method. They only numerically investigated the case of the extreme value index with $\gamma$=-0.2. In addition, the PWM and the ML method do not work well for an extreme value index of $\gamma$<-0.5 (Coles, 2011, section 3.31) as it is for the TED with $\gamma$=-1. The ML method has a large bias (Hosking et al. 1985, Fig. 2) and the PWM has a small efficiency compared with the ML method (Hosking et al. 1985, Fig. 4). The authors do not present an argument why the moment method could work better for small extreme value indexes with $\gamma$≤-0.25.

Besides, Pisarenko et al. apply the ML method to estimate the Kolmogorov distance for the GEVD (their Eq. (26)), which is not consistent to their point estimation with the moment method.

The Kolmogorov distance is used by the authors to detect the optimal threshold of the POT analysis. This distance is the test statistics of the Kolmogorov-Smirnov goodness-of-fit test. The idea of applying a goodness-of-fit statistics for the threshold selection is not new. For example, the procedure by Goegebeur et al. (2008) includes a goodness-of-fit statistics, is based on a stringent theory and is validated by numerical investigation of different situations. Pisarenko et al. do not provide any such validation.

The POT analysis is generally used in extreme value statistics for estimating an extreme value index because the sample size is normally larger than that of the block maxima. Further, the GPD uses only two parameters. Both result in a smaller estimation error of the extreme value index than the estimation with the block maxima. As mentioned before, the authors estimate the parameters of the GPD in the POT analysis by the ML method. In case of a small sample size, the ML method does not work well for a small extreme value index of $\gamma \leq -0.25$ (Hüsler et al., 2011, Fig. 3 and 5). In addition, the equations of the ML methods do not solve every sample in case of $\gamma < -0.5$ according to Grimshaw (1993). The estimation method of Hüsler et al. (2011 and 2014) has not such a limitation, and is hence recommended for a small extreme value index. It is recalled that the extreme value index of the important TED is $\gamma = -1$.

A further crucial point is the error quantification in the estimation procedure of Pisarenko et al., which includes a bootstrap, an averaging, and a Monte Carlo simulation for applying GEVD and the GPD. Finally, the authors use the average of the estimated parameters for different block lengths in the case of a GEVD and different thresholds in the case of a GPD. I do not see the advantage of such an averaging. On the contrary, if simple point estimation is used, then the asymptotic variance-covariance matrix of the estimation method can be used for quantifying the estimation error (Hüsler et al., 2011, ch. 2.1.). The authors also quantify the standard deviation of the estimation error in a different way. They apply a reshuffling according to the bootstrap method and use 100 reshufflings, which seems to be too few according to DasGupta (2008). The standard deviations (standard errors) shown in their Figures 2 to 8 are computed using this procedure. Confusingly, there are different standard deviations for the lower and the upper lines in these figures (e.g. Fig. 6, T=7), although there is only one standard deviation. Furthermore, Pisarenko et al. compute the standard deviation and the mean squared error (MSE) by using Monte Carlo simulations with 500 generated samples of a GEVD and of a GPD in case of the POT analysis. The results for the standard deviations by the bootstrap and the Monte Carlo simulation differ extremely. For example, for the GEVD of the Havard catalog we have an MSE($\gamma$)=0.047 for T=80 days, which means a standard deviation of Std($\gamma$)≈0.21. But the standard deviation in the corresponding Fig. 2 of the authors has the value Std($\gamma$)=0.02-0.025 for T=75 days, which is only a small fraction of 0.21. I strongly reject their interpretation that '*This is not surprising since the latter gives only the scatter conditional to the same unique data sample*', because the bootstrap method is specifically used the quantifying the error distribution and its corresponding standard error (DasGupta, 2008). The large difference might be an indicator for the fact that the approximating the distribution of extreme magnitudes by GEVD and GPD does not work well.

This leads again to the main problem of the poor convergence speed in applying GPD and GEVD to the earthquake magnitudes. As aforementioned, the GEVD and the corresponding estimation methods do not work well for the TED (Raschke, 2012). The upper bound magnitude should be estimated by the methods of Pisarenko et al. (1996), Kijko and Graham (1998), Hannon and Dahiya (1999) or Raschke (2012), where the quantile of a maximum magnitude of a defined time period can be computed by the inverse of Eq. (7). The limitation of the extreme value theory and statistics also applies to other distribution models for magnitudes because they are very similar to the ED according to the Gutenberg-Richter relation.

## 4. Is a homogeneous Poisson process needed?

Finally, I want to point out that earthquake data do not need to occur as a homogeneous Poisson process for applying extreme value theory and statistics. Pisarenko et al. decluster the earthquake catalog and use only the main events to provide a homogeneous Poisson process with independent magnitudes similar to e.g. Zöller et al. (2014). But a homogeneous Poisson process is not a necessary condition for applying GEVD and GPD. A homogeneous Poisson process would imply that the number of events with a magnitude over a defined threshold would be Poisson distributed (Johnson et al. 1994, p. 553). But the GEVD works asymptotically for different distributions of excess numbers. A simple example is the maxima of an exponentially-distributed random variable with a fixed, not Poisson distributed block (sample) size that converges quickly to a GEVD (Beirlant et al., 2004, Fig. 2.9). Leadbetter et al. (1983, Part II) and Falk et al. (2011, Part III) give further extensions. The GEVD can also be applied for random variables with serial correlation under certain conditions. The extremal index, an additional parameter, compensates the influence of the serial correlation.

In addition, it has been shown that the parameters of the GPD can be estimated in a POT analysis even if there is serial correlation between the members of excess clusters (Raschke, 2013a). Therein, the estimation error can be quantified with the Jack-knife method. The crucial point of the earthquake magnitudes is the poor convergence of their tail to the GPD, not the earthquake process in time.

## 5. Conclusions

In this comment, I have discussed important aspects of the research of Pisarenko et al. from the perspective of extreme value statistics and theory. In summary, I advise against using the procedures as applied by Pisarenko et al.. GPD and GEVD work only well for the extremes of TED or GTED when the block size is very large and/or the threshold is very close to the upper bound magnitude. The crucial point of the earthquake magnitudes is the poor convergence of their upper tail to the GPD. Therefore, the classical methods for estimating the upper bound of the TED should be applied as shown by Pisarenko et al. (1996), Kijko and Graham (1998), Hannon and Dahiya (1999) and Raschke (2012). The appropriateness of these methods for other distribution models such as the GTED should be examined in further research. Corresponding parameters of the maximal magnitudes such as the quantile of random block maxima of a defined time period can be computed by the inverse of Eq. (7) in all cases.

# Appendix

Tab. A1: Applied symbols and notations

| in comments | in Pisarenko et al. | Explanation |
|---|---|---|
| $\beta$ | $\beta$ | Scale parameter of the TED (Eq. (1)) |
| $\gamma$ | $\xi, \gamma, \zeta$ | Extreme value index of GPD and GEVD |
| $\sigma^*$ | s | Scale parameter of GPD, our Eq. (3) |
| $\sigma$ | $\sigma$ | Scale parameter of GEVD, our Eq. (5) |
| $\mu$ | $\mu$ | Location parameter of GEVD, our Eq. (5) |
| $m_{min}$ | m | Defined or selected lower bound magnitude of the TED, parameter in Eq. (1) |
| $m_{max}$ | $m_{max}$ | Upper bound magnitude of the TED (parameter in Eq. (1)) and also upper bound of GPD and GEVD for $\gamma<0$ in case of earthquake magnitudes |
| n | n | Sample size, block size; can also be the average block size for a defined time period in Eq. (7) |
| $n_{thresholds}$ | $\lambda T$ | Sample size of the excess; can also be the average sample size |
| x | x | Scale of the real numbers |
| $x_{threshold}$ | H | Threshold in our Eq. (2) |
| X | | Random variable |
| Y | X-H | Random variable, excess  Y=X- $x_{threshold}$, our Eq. (2) |
| Z | $M_n$, $M_T$ | Block maximum, maximum magnitude of a time period, our Eq. (4) |
| F(x) | F(x) | Cumulative distribution function (CDF) of a random variable X; expresses the probability of non-exceedance X≤x |
| G(x) | $\Phi(x)$ | CDF of block maximum Z, CDF of the GEVD, our Eq. (5-8) |
| H(x) | $F_H(x)$ | CDF of a random excess Y, CDF of the GPD |
| Pr{a │ b} | | Probability of a under the condition of b |